**Title:**

Amorphous interface layer in thin graphite films grown on the carbon face of SiC

**Authors:**

R. Colby[1,2], M.L. Bolen[2,3], M.A. Capano[2,3], E.A. Stach[4]

**Affiliations:**

1. School of Materials Engineering, Purdue University
2. Birck Nanotechnology Center
3. School of Electrical and Computer Engineering, Purdue University
4. Center for Functional Nanomaterials, Brookhaven National Laboratory

**Abstract:**

Cross-sectional transmission electron microscopy (TEM) is used to characterize an amorphous layer observed at the interface in graphite and graphene films grown via thermal decomposition of C-face *4H*-SiC. The amorphous layer does not to cover the entire interface, but uniform contiguous regions span microns of cross-sectional interface. Annular dark field scanning transmission electron microscopy (ADF-STEM) images and electron energy loss spectroscopy (EELS) demonstrate that the amorphous layer is a carbon-rich composition of Si/C. The amorphous layer is clearly observed in samples grown at 1600°C for a range of growth pressures in argon, but not at 1500°C, suggesting a temperature-dependent formation mechanism.

**Text:**

The formation of epitaxial graphene and graphite by thermal decomposition of SiC has been studied for over three decades, but has seen a dramatic revitalization since the discovery of the electric-field effect in graphene in 2004 [1]. The majority of the work based on SiC has focused on the silicon-face SiC-$(0001)$ surface, typically in a vacuum growth environment [2,3]. The carbon-face SiC-$(000\bar{1})$ surface has received less attention, but typically results in thin graphite films, frequently referred to as multi-layer graphene due to observations of graphene-like band structures [4,5,6]. As a result, the SiC/graphite interface is typically buried beneath several layers of graphene, and the structure is not immediately interpretable by the commonly applied surface sensitive techniques. The coupling of the first graphene layer and the C-face SiC substrate has thus far been described as both strong and weak, with a reported interface spacing varying from ~1.6 Å–3.2 Å [6,7,8]. Most techniques for assessing the graphite film thickness require a model for the SiC/graphite interface, the structure of which remains likewise ambiguous, and is not necessarily uniform across an entire sample [9,10,11]. Cross-sectional transmission electron microscopy (TEM) offers a more direct means of examining the film thickness and the SiC/graphite interface. The relative thicknesses of the graphite films can be determined locally, without the need for additional (often unknown) material properties, often without making assumptions as to the nature of the SiC/graphite interface. The latter benefit should become particularly apparent; this letter will report on the observation of an unexpected carbon-rich amorphous interface layer for graphite films grown on C-face *4H*-SiC over a range of pressures.

Nominally on-axis *4H*-SiC substrates obtained from Cree were cleaned in solvent and acid baths then etched *in situ* with $H_2$ gas in an Epigress VP508 hot-walled reactor to remove surface damage. The graphene and graphite thin films were formed over 10 minutes of thermal decomposition in vacuum (low-$10^{-5}$ mbar), or in an argon environment, at 1500°C and 1600°C [12]. Cross-sectional TEM samples were prepared using a focused ion beam (FIB) liftout method on an FEI Nova FIB/SEM, equipped with a Klöcke nanomanipulator. Protective layers of Pt/C were deposited locally in the area of interest, initially with the electron beam to avoid surface damage. TEM, scanning transmission electron microscopy (STEM), and energy electron loss spectroscopy (EELS) were obtained with an FEI Titan 80-300 operating at 300 kV, equipped with a Gatan imaging filter, with images zero-loss filtered to improve contrast. An FEI Tecnai 20 operated at 80 kV was used to check for potential electron-beam-induced effects. Cross-sections were initially left fairly thick (greater than ~100 nm) to mitigate possible ion beam damage effects, owing to the sensitivity of graphite films [13,14]. While the effects of beam damage were certainly apparent, it was possible to obtain meaningful images, even for cross-sections thinned to less than 60 nm.

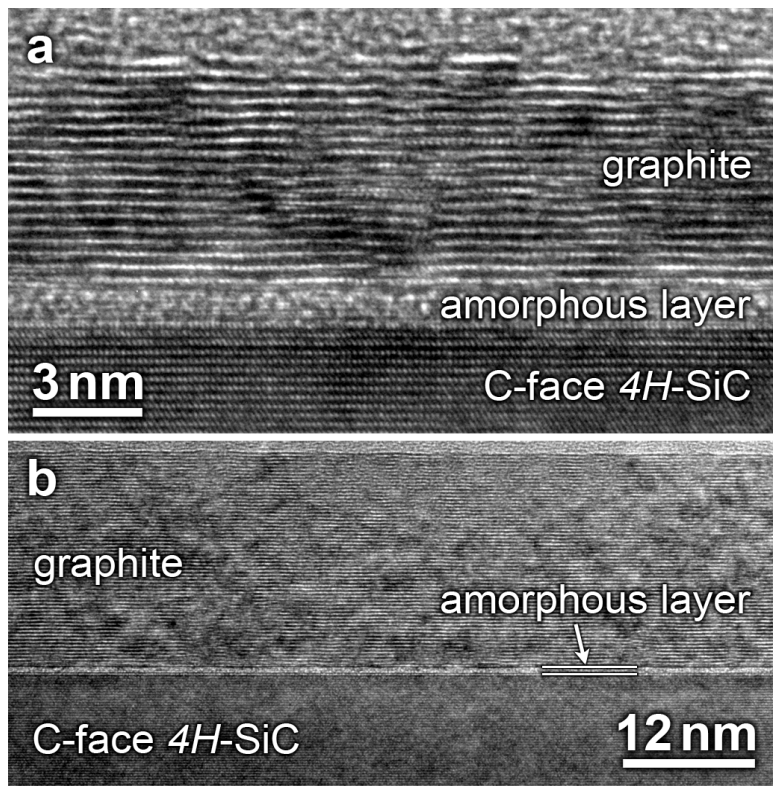

Figure 1: (a) High resolution TEM image of a typical region of the C-face *4H*-SiC and graphite film interface containing an amorphous intermediate layer, grown at 1600°C and 50 mbar of Ar. The graphite film is ~6.8 nm thick, while the amorphous layer is ~10 Å (the exact thickness is obscured by delocalization of the adjacent crystalline layers). (b) TEM image of another typical region grown at 1600°C in vacuum, with a far thicker graphite film (~21 nm), yet a thinner amorphous layer (~5 Å). The slightly mottled contrast is a product of ion and electron beam damage, especially prominent in the extremely beam-damage sensitive graphite. Both images were acquired in the SiC-$\langle 1\bar{1}00\rangle$ direction.

An amorphous layer was observed at the SiC/graphite interface in regions of all examined samples grown at 1600°C, including samples grown at pressures of $10^{-5}$ mbar (Fig.1b), 1 mbar of Ar, and 50 mbar of Ar (Fig.1a). In each region where the amorphous layer was observed, it was locally uniform in thickness, usually persisting with no discernible variation in thickness across several microns of interface. The thickness of the amorphous layer varied from sample to sample, ranging from ~5–12 Å. Graphite film thicknesses varied between samples, and by region within samples, over a range of ~4–21 nm. Likewise, there was no clear correlation between amorphous layer thickness and either the graphite film thickness, or growth pressure.

The amorphous layer was more clearly defined in annular dark-field (ADF) STEM images, which are conveniently bereft of the delocalization effects seen in high-resolution TEM images (Fig.2a). The (0001) planes were clearly resolved in both the substrate and the graphite, and just as clearly absent in the amorphous layer. Additionally, ADF-STEM can be collected at higher scattering angles to obtain Z-like contrast (Fig.2b). The intensity of the amorphous layer, and thus the Z-number, was in between that of the SiC substrate and the graphite film; in the absence of an unexpected contaminant, this suggested a composition of Si/C in between the SiC substrate and the pure carbon graphite. As the microscope employed was limited to annular detector outer angle of ~70 mrad, the signal obtained at angles suitable for Z-like contrast was quite low. Figure 2b has, therefore, been averaged along $\langle 0001 \rangle$ to more clearly illustrate the layers' relative intensities.

EELS was utilized to more directly confirm the composition of the amorphous layer. EELS was collected with the focused STEM probe at several points along the interface from within the SiC substrate, the amorphous layer, and the graphite film. The electron beam was sufficiently localized to isolate the signals from the individual layers. As expected, both the Si-$L_{3,2}$ and C-K edges were apparent for the SiC substrate, while only the C-K edge emerged for the graphite film (Fig.2c). The amorphous layer contained both silicon and carbon edges. No other edges were observed in any of the layers in proximity to the amorphous layer. By comparison with the nominally 1:1 stoichiometric SiC substrate, the relative Si:C ratio was found to be ~1:4 within the amorphous layer shown in figure 2(a), consistent with the relative intensities observed in STEM images (EELS quantification details included in *supplementary information*). It is possible that the precise Si:C ratio varied from sample to sample, or even through the thickness of the amorphous layer, but the sensitivity of these samples to beam damage limited further analysis. Likewise, the energy resolution and signal to noise were insufficient to allow a more quantitative assessment of the apparent shifts and near edge structure differences for the relevant edges.

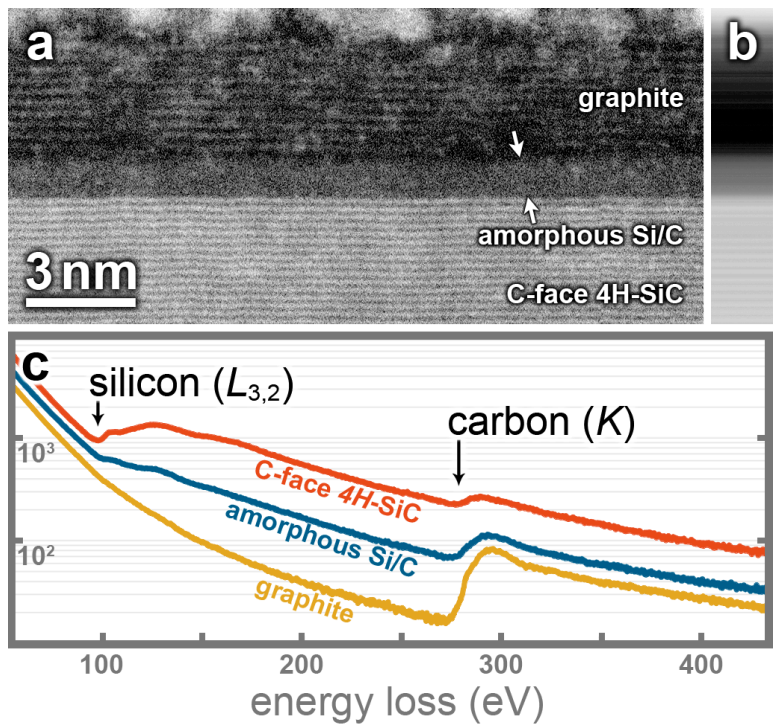

**Figure 2:** (a) ADF-STEM image of a typical region of a graphite film grown on C-face *4H*-SiC at 1600°C and 1 mbar of Ar containing an amorphous intermediate layer. Imaged from the $\langle 1\bar{1}00 \rangle$ direction, the (0001) lattice planes of both the substrate and graphite film are visible, but absent in the amorphous layer. (b) Higher angle ADF-STEM image, with Z-like contrast, averaged across $\langle 0001 \rangle$ to reduce noise (the width of the included image is arbitrary). (c) EELS collected locally (using the STEM probe) from each of the three regions shows that the amorphous layer contains both Si and C, with a ~1:4 Si:C, in rough agreement with conclusions drawn from the contrast in (b). The spectra are vertically offset from each other for clarity, along a log-scale vertical axis in units of arbitrary intensity.

It should be specifically noted that the amorphous layer was not an artifact of sample preparation; in the very least, the formation of a silicon-containing layer above the only potential Si source (the SiC) would have been quite improbable in the top-down milling geometry used in the FIB. More direct evidence was fortuitously found in cross-sections of the wrinkles or ridges that are known to populate graphite films on C-face SiC [15,16,17], prior to any sample preparation (Fig. 3). The amorphous Si/C layer was observed spanning the gap beneath several such locally delaminated graphite ridges. The contrast of the amorphous layer was clearly distinct from the redeposited amorphous carbon material underneath the graphite ridge (confirmed with EELS for a similar ridge). The more exposed portion of the amorphous Si/C layer beneath the ridge even appears to have suffered some ion beam damage, firmly implying that it was present prior to cross-section preparation.

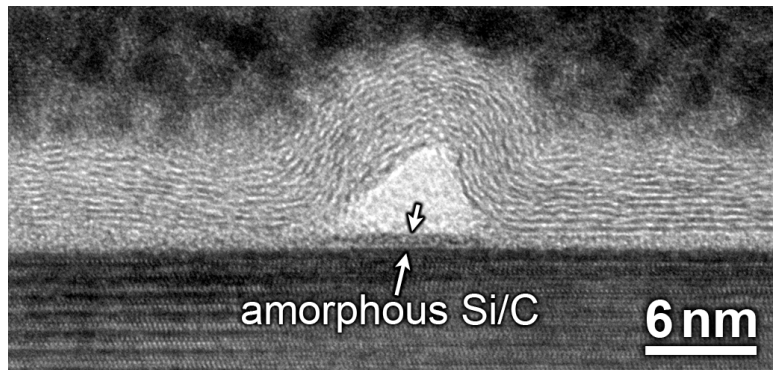

Figure 3: Cross-sectional TEM image of one of the characteristic ridges observed in graphite grown on C-face SiC. The amorphous Si/C layer continues beneath the ridge, and is distinct from the amorphous carbon material that has collected beneath the ridge as a result of sample preparation. Image acquired in the SiC-$\langle 11\bar{2}0 \rangle$ direction.

Few-layer graphene films grown under similar conditions on Si-face SiC substrates were not observed to contain any such amorphous layer, in agreement with previous publications. Graphite films grown on C-face SiC at a lower temperature of 1500°C could not be definitively ascribed as containing an amorphous layer either, though the interface was indistinct in many places and there did appear to be a larger spacing at the SiC/graphene interface than for the comparable Si-face samples (this latter point has been noted previously [6,18]). Therefore, some threshold for the formation of the amorphous layer seems to have existed, being either dependant upon temperature or temperature-related processing conditions. In the bulk, there are more than twice as many carbon atoms in graphite (by volume) than in *4H*-SiC. It seems plausible that a carbon-rich layer of Si/C would have accumulated prior to the formation of a stable graphene layer. The amorphous layer was reminiscent of the amorphous intergranular films frequently observed in liquid-phase sintered bulk SiC [19,20], usually attributed to residual additives or contaminants diffusing towards the SiC grain boundaries (e.g. oxygen). The excess carbon near the SiC/graphite interface could have served as an analogous "contamination." As for the case of intergranular films, the formation of an amorphous wetting layer may have mitigated the expected interfacial strain: the in-plane lattice mismatch between graphite and SiC is ~20% at room temperature, and in-plane thermal expansion mismatch ~60−80% [16,21,22]. However, the kinetics of graphene growth on C-face SiC are not yet well enough understood to postulate a more precise mechanism, or to be sure as to whether the resulting amorphous layer would have formed during growth, or during the cool-down to room temperature.

To summarize, an amorphous Si/C layer has been observed at the interface between C-face 4H-SiC and graphite films grown by thermal decomposition at 1600°C, from vacuum to 50 mbar of Ar. The layer was frequently uniform across microns of cross-sectional interface, as observed by TEM, with contiguous segments each fairly uniform in thickness. There were variations between regions and samples with no clear correlation to growth pressure or the overlaying graphite film thickness, though there seems to have been a growth temperature-related threshold for its emergence. EELS confirmed that the layer was a carbon-rich ~$Si_1/C_4$. Without a larger-scale means of detecting the amorphous layer under the surface, it has not been possible to investigate the potential consequences of the amorphous layer: but such effects might be expected, for instance, on the mobility of the graphene or graphite films, and therefore on any devices fabricated on such films.

**Supplementary Information:**

The relative Si/C composition of the amorphous layer was estimated by comparison with the nominally stoichiometric SiC substrate. The pre-edge backgrounds were each independently fit to a power law model, as is typical for EELS, over a 20 eV window ending 15 eV prior to each edge. The post-edge intensities were integrated over a 50 eV window, offset 5 eV forwards of each edge to avoid differences in near-edge structure of the layers. The relative carbon to silicon ratio was found to be ~1:4 within the amorphous layer by assuming a 1:1 silicon to carbon ratio within the substrate and scaling the integrated intensities accordingly, obviating the necessity of using a model for the relevant cross-sections. However, the standard quantification algorithm included with Digital Micrograph (Gatan, Inc.) electron microscopy software, using a Hartree-Slater model, did yield a ~1:1 ratio in the substrate and similar ~1:4 ratios in the amorphous interface layer.